\documentclass[pra,onecolumn,preprintnumbers,amsmath,amssymb,nofootinbib]{revtex4}
\usepackage{graphicx}
\graphicspath{ {images/} }
\usepackage{listings}
\usepackage{color}
\usepackage{amsmath}
\usepackage{amsthm}
\usepackage{physics}
\usepackage{qcircuit}
 
\definecolor{codegreen}{rgb}{0,0.6,0}
\definecolor{codegray}{rgb}{0.5,0.5,0.5}
\definecolor{codepurple}{rgb}{0.58,0,0.82}
\definecolor{backcolor}{rgb}{0.95,0.95,0.92}
 
\lstdefinestyle{mystyle}{
    backgroundcolor=\color{backcolor},   
    commentstyle=\color{codegray},
    stringstyle=\color{codepurple},
    basicstyle=\footnotesize\ttfamily,
    breakatwhitespace=false,         
    breaklines=true,                 
    captionpos=b,                    
    keepspaces=true,                 
    showspaces=false,                
    showstringspaces=false,
    showtabs=false,                  
    tabsize=2
}
\lstnewenvironment{mylstlisting}{\lstset{style=mystyle}}{}

\lstdefinestyle{mycodestyle}{
    backgroundcolor=\color{backcolor},   
    commentstyle=\color{codegreen},
    keywordstyle=\color{magenta},
    numberstyle=\tiny\color{codegray},
    stringstyle=\color{codepurple},
    basicstyle=\footnotesize\ttfamily,
    breakatwhitespace=false,         
    breaklines=true,                 
    captionpos=b,                    
    keepspaces=true,                 
    numbers=left,                    
    numbersep=5pt,                  
    showspaces=false,                
    showstringspaces=false,
    showtabs=false,                  
    tabsize=2
}

\begin{document}
\title{Towards An Implementation of the Subset-sum Problem on IBM's Quantum Experience}

\author{David Gunter}
\thanks{Corresponding author: dog@lanl.gov; LA-UR-18-28220}
\author{Toks Adedoyin}
\affiliation{Future Architectures Team; Computer, Computational, and Statistical Sciences Division; Los Alamos National Laboratory, Los Alamos, NM 87545, USA}

\date{November 2017}

\begin{abstract}
   In seeking out an algorithm to test out the capability of IBM's Quantum Experience quantum computer,
   we were given a review paper\cite{Bernstein2013} covering various algorithms for solving the
   subset-sum problem, including both classical and quantum algorithms. The paper went on to
   present a novel algorithm that beat the previous best algorithm known at the time. The complex
   nature of the algorithm made it difficult to see a path for implementation on the Quantum
   Experience machine and the exponential cost--only slightly better than the best classical
   algorithm--left us looking for a different approach for solving this problem. 
   
   We present here a new quantum algorithm for solving the subset-sum problem that for many cases
   should lead to $\mathcal{O}(\text{poly}(n))$-time to solution. The work is reminiscent of the
   verification procedure used in a polynomial-time algorithm for the quantum Arthur-Merlin games
   presented in \cite{Marriott2005}, where the use of a quantum binary search to find a maximum
   eigenvalue in the final output stage has been adapted to the subset-sum problem as in
   \cite{Daskin2017}.
\end{abstract}
\maketitle

\section{Introduction}
\subsection{Subset-Sum}
The well-known subset-sum problem is easily stated as follows. Given a set of $n$ integers $X_n = \{x_1, x_2, \ldots, x_n\}$, and a target integer $s$, the subset-sum problem is to determine whether the sum of any subset $X_m \subseteq X_n$ is equal to $s$.

Let $I_m$ be a subset of the indices $\{1,2,\ldots,n\}$ containing $m \leq n$ elements, then what we want to determine is if there is a subset $X_m$ that satisfies
\begin{equation}
        s = \sum_{i \in I_m} x_i. \nonumber 
\end{equation}

The subset-sum problem has been proven to be NP-complete. Therefore, any classical polynomial-time algorithm that would solve the problem would violate the conjecture $P \neq NP$. As such, the best classical algorithm discovered to date solves the problem in time $t \propto 2^{0.291 n}$.

Recently, the first quantum algorithm for solving the subset-sum problem that beats the best classical algorithm was proposed by Berstein \textit{et al}\cite{Bernstein2013}. They developed a faster exponential-time quantum algorithm that solves the problem in time $t \propto 2^{0.241 n}$.

We present in this paper a quantum algorithm that for many cases solves the subset-sum problem in polynomial time, giving exponential speedup over the classical solution.

\section{Quantum Algorithm For Solving The Subset-Sum Problem}

The basic idea is to encode the $2^n$ possible subset sums as the eigen-phases ($\varphi$) of a diagonal unitary matrix $U$. With $U$ suitably defined, quantum phase estimation (QPE) is used to determine the values of the subset sums (the eigenvalues) and the associated eigenvectors, as per the detailed discussion in Nielson \& Chuang\cite{NielsenChuang}\footnote{Starting on page 221.}. 

Next, amplitude amplification\cite{NielsenChuang}\footnote{Starting on page 250.} is used to to single out those sums less than or equal to $s$ and to eliminate the states with sums greater than $s$. Left with only possible-good states, amplitude amplification is used one final time to find the maximum phase representing the solution to the subset-sum problem.  These steps are explained in detail in the following subsections.

In order to make use of QPE we will require two different multi-qubit registers. The first is a $t$-qubit register, with all qubits initialized to $\ket{\mathbf{0}}$. $t$ is chosen based on the number of digits of accuracy we would like to achieve for the eigenvalues (subset-sums). The second register contains $n$ qubits and is used to store the final eigenvector representing the elements of $X_n$ in the solution to the problem. It will become clear in what follows that we must have $t \geq n$.

\subsection{Quantum Phase Estimation: Subset-sums As The Phases Of A Unitary Matrix}

The idea behind quantum phase estimation is that we have a unitary matrix $U$ and an eigenvector $\ket{u}$ with a corresponding eigenvalue $e^{2\pi i \varphi}$, and we want to find this eigenvalue. This means finding the value of the phase variable $\varphi$. Since we can only determine the overall phase in the exponential to a value modulo $2\pi$, we can only produce a value of $\varphi$ in the range $0 \leq \varphi < 1$. Our goal is to create a unitary matrix $U$ in such a way that the set of of possible eigenvalues contain---in the phase variables---the set of all possible subset-sum values from our set $X_n$.  In order to do this, however, we will have to scale the elements of $X_n$ in such a way that the largest possible sum value is less than unity, i.e. $\sum_{j=0}^{n-1} x_j < 1$. We choose a scaling such that $\sum_{j=0}^{n-1} \bar{x}_j \leq 0.5.$ where the $\bar{x}_j$ now denotes our scaled values of the original set $X_n$. 

We define our unitary matrix $U$ for use in the QPE step as a Kronecker product of a set of rotation matrices $R_j, j \in [0,n),$ where

\[ R_j = \left( \begin{matrix}
                        1 & 0 \\
                        0 & e^{2\pi i \bar{x}_j}
         \end{matrix} \right). \]
That is,

\[ U = \left( R_{n-1} \otimes R_{n-2} \otimes \ldots \otimes R_0 \right). \]
It is easy to verify that $U$ is a purely diagonal matrix with elements given as

\[ \left( 1, e^{2\pi i \bar{x}_0}, e^{2\pi i \bar{x}_1}, e^{2\pi i (\bar{x}_0 + \bar{x}_1)}, \ldots, e^{2\pi i (\bar{x}_0 + \bar{x}_1 + \bar{x}_2 + \ldots + \bar{x}_{n-1})} \right). \]
The elements (eigenvalues) of $U$ comprise the set of phase values representing every possible subset sum for the scaled set $\bar{X}_n$. Notice that the $R_j$ operators can be implemented by simple $\pi/8$ or \textit{phase-shift} gates.

Observe that 

\[ U^2 \ket{u} = U e^{2\pi i \varphi} \ket{u} = e^{2\pi i \varphi} U \ket{u} = e^{2\pi i 2\varphi} \ket{u}, \]
or in general that

\[ U^{2^j} \ket{u} = e^{2\pi i 2^j \varphi} \ket{u}. \]

We now have everything we need to implement the QPE algorithm as described in \cite{NielsenChuang}. We choose as the initial state of our registers

\[ \ket{\psi_0} = \ket{\mathbf{0}} \frac{1}{\sqrt{2^n}} \sum_{j=0}^{2^n-1} \ket{j}. \]
That is, Register 1 has all $t$ qubits in the state $\ket{\mathbf{0}}$ as discussed previously, and the $n$ qubits of Register 2 are in a superposition representing the total enumerations of all possible subsets $X_n$. This initial state can easily be generated by a combination of identity and Hadamard gates, i.e.

\begin{equation}
        \ket{\psi_0} = \left( I^{\otimes t} \otimes H^{\otimes n} \right) \ket{\mathbf{0}} \ket{\mathbf{0}}. \label{psi0}
\end{equation}

\begin{figure}
\begin{equation*}
\Qcircuit @C=1.0em @R=.7em {
  &&& \lstick{\ket{0}} &  \gate{H}       & \qw            & \qw            & \qw            & \qw & \ldots & & \ctrl{9}           & \qw \\
  &&& \lstick{}        &                 &                &                &                &     &        & &                    & \\
  &&& \lstick{}        & \vdots          &                &                &                &     &        & &                    & \\
  &&& \lstick{}        &                 &                &                &                &     &        & &                    & \\
  &&& \lstick{\ket{0}} &  \gate{H}       & \qw            & \qw            & \ctrl{5}       & \qw & \ldots & & \qw                & \qw \\
  &&& \lstick{\ket{0}} &  \gate{H}       & \qw            & \ctrl{4}       & \qw            & \qw & \ldots & & \qw                & \qw \\
  &&& \lstick{\ket{0}} &  \gate{H}       & \ctrl{3}       & \qw            & \qw            & \qw & \ldots & & \qw                & \qw \\
  &&& \lstick{}        &                 &                &                &                &     &        & &                    & \\
  &&& \lstick{}        &                 &                &                &                &     &        & &                    & \\
  &&& \lstick{\ket{u}} &  \qw {/^n}      & \gate{U^{2^0}} & \gate{U^{2^1}} & \gate{U^{2^2}} & \qw & \ldots & & \gate{U^{2^{t-1}}} & \qw  \\
  &&& \lstick{}        &                 &                &                &                &     &        & &                    & \\
  &&& \lstick{}        &                 &                &                &                &     &        & &                    & 
  {\inputgroupv{1}{7}{1.2em}{4.0em}{\rotatebox{90}{\text{Register 1}}}} \\
  {\inputgroupv{9}{11}{0.4em}{1.6em}{\rotatebox{90}{Register 2}}} \\
}
\end{equation*}
\caption{The circuit diagram for the quantum phase estimation algorithm}
\label{QPEDetail}
\end{figure}
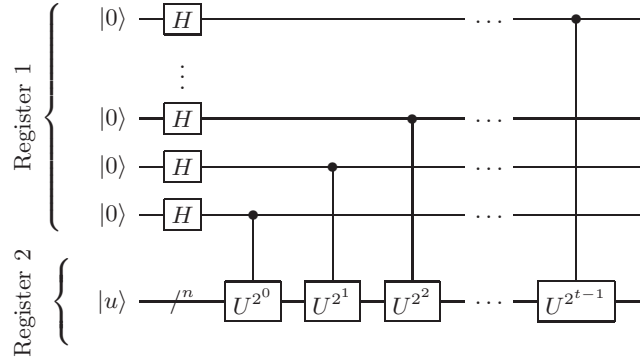

We start by applying Hadamard gates to each qubit of Register 1, leaving them in the equal superposition of the states $\ket{\mathbf{0}}$ and $\ket{\mathbf{1}}$. We then apply a total of $t$ Control-$U^{2^j}$ gates, $0 \leq j < t$, where each $U^{2^j}$ operates on Register 2 depending on the state of the $j^{th}$ qubit of Register 1. This is illustrated in Fig. \ref{QPEDetail}.

Here is the action of any one of these gates on the state $\frac{1}{\sqrt{2}} (\ket{0} + \ket{1}) \ket{u}$:

\begin{align}
  \text{Control}-U^{2^j} \left[ \frac{1}{\sqrt{2}} (\ket{0} + \ket{1}) \ket{u} \right] &= \frac{1}{\sqrt{2}} (\ket{0} \ket{u} + \ket{1} U^{2^j} \ket{u}) \nonumber \\
&= \frac{1}{\sqrt{2}} (\ket{0} \ket{u} + \ket{1} e^{2\pi i 2^j \bar{x}_j} \ket{u}) \nonumber \\
&= \frac{1}{\sqrt{2}} (\ket{0} \ket{u} + e^{2\pi i 2^j \bar{x}_j} \ket{1} \ket{u}) \nonumber \\
&= \frac{1}{\sqrt{2}} (\ket{0} + e^{2\pi i 2^j \bar{x}_j} \ket{1}) \ket{u}) \nonumber
\end{align}

After all $t$ Control-$U^{2^j}$ gates have been applied in this manner, the state of Register 1 can be described by

\begin{align}
\frac{1}{\sqrt{2^t}} ( \ket{0} + e^{2\pi i 2^{t-1} \varphi} \ket{1} ) ( \ket{0} + e^{2\pi i 2^{t-2} \varphi} \ket{1} ) \ldots ( \ket{0} + e^{2\pi i 2^0 \varphi} \ket{1} ) 
= \frac{1}{\sqrt{2^t}} \sum_{j=0}^{2^t - 1} e^{2\pi i j \varphi} \ket{j} . \label{postCUstate}
\end{align}
All the information about the phases (the subset sums) is contained in Register 1. Before proceeding, let us write $\varphi = \tilde{\varphi} + \delta 2^{-t}$, where $\tilde{\varphi}$ is the first $t$ bits in the binary expansion $0.\varphi_1 \varphi_2 \ldots \varphi_t$ and the $\delta$-term represents the remaining bits. If the expansion were exactly $t$ bits, i.e. $\delta = 0$, this would mean

\begin{align}
  e^{2\pi i \varphi} &= e^{2\pi i 0.\varphi_1 \varphi_2 \ldots \varphi_t}, \nonumber \\
  e^{4\pi i \varphi} &= e^{2\pi i \varphi_1 . \varphi_2 \ldots \varphi_t} = e^{2\pi i \varphi_1} e^{2\pi i 0.\varphi_2 \ldots \varphi_t} = 
          e^{2\pi i 0.\varphi_2 \ldots \varphi_t}, \nonumber \\
  \vdots \nonumber \\
  e^{2^j \pi i \varphi} &= e^{2\pi i .\varphi_j \ldots \varphi_t}. \nonumber
\end{align}
We can then write the post Control-$U^{2^j}$ state, Eq. \ref{postCUstate}, as
\begin{align}
\frac{1}{\sqrt{2^t}} ( \ket{0} + e^{2\pi i 0.\varphi_1 \varphi_2 \ldots \varphi_t} \ket{1} ) \otimes \ldots \otimes ( \ket{0} + e^{2\pi i 0.\varphi_t} \ket{1} ), 
\end{align}
which is the Fourier transform of a basis state
\[ \ket{\varphi_1 \varphi_2 \ldots \varphi_t} = \ket{2^t\varphi}. \]
We only need apply a $t$-qubit inverse quantum Fourier transform (QFT$^{-1}$) to Register 1 which converts the qubits to the state $\ket{\varphi_1} \otimes \ket{\varphi_2} \otimes \ldots \otimes \ket{\varphi_t}$, where each $\varphi_j (1 \leq j \leq t)$ is an estimated bit equal to 0 or 1. 

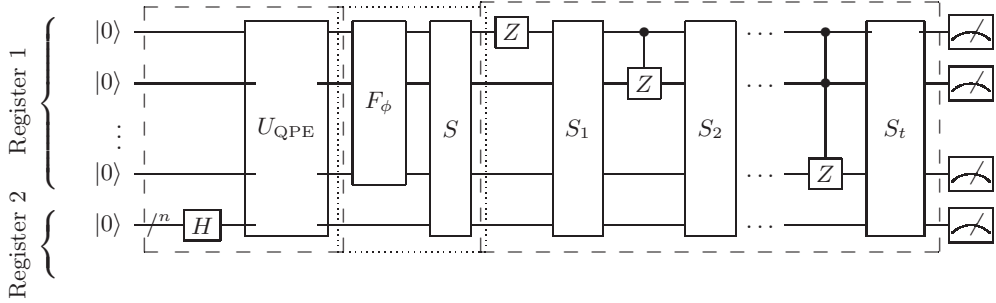
\begin{figure}
\begin{equation*}
\Qcircuit @C=1em @R=.7em {
   &&&  \lstick{\ket{0}} &   \qw   &   \qw    & \multigate{4}{U_{\text{QPE}}} & \multigate{3}{F_{\phi}} & \multigate{4}{S} & \gate{Z} & \multigate{4}{S_1} & \ctrl{1} & \multigate{4}{S_2} & \ldots & & \ctrl{3} & \multigate{4}{S_t} & \meter \\
   &&&  \lstick{\ket{0}} &   \qw   &   \qw    & \ghost{U_{pe}}        & \ghost{F_{\phi}}        & \ghost{S}        & \qw      & \ghost{S_1}        & \gate{Z} & \ghost{S_2}        & \ldots & & \ctrl{2} & \ghost{S_m}        & \meter \\
   &&&  \lstick{\vdots}  &         &          & \pureghost{U_{pe}}    & \pureghost{F_{\phi}}    & \pureghost{S}    &          & \pureghost{S_1}    &          & \pureghost{S_2}    &        & &          & \pureghost{S_m}    &        \\
   &&&  \lstick{\ket{0}} &   \qw   &   \qw    & \ghost{U_{pe}}        & \ghost{F_{\phi}}        & \ghost{S}        & \qw      & \ghost{S_1}        & \qw      & \ghost{S_2}        & \ldots & & \gate{Z} & \ghost{S_m}        & \meter \\
   &&&  \lstick{\ket{0}} & {/^n} \qw & \gate{H} & \ghost{U_{pe}}        &  \qw                    & \ghost{S}        & \qw      & \ghost{S_1}        & \qw      & \ghost{S_2}        & \ldots & & \qw      & \ghost{S_m}        & \meter 
        \gategroup{1}{5}{5}{7}{1.2em}{--} \gategroup{1}{8}{5}{9}{1.2em}{..} \gategroup{1}{10}{5}{17}{1.2em}{--} \\
  {\inputgroupv{1}{4}{1.1em}{2.8em}{\rotatebox{90}{\text{Register 1}}}} \\
  {\inputgroupv{5}{6}{1.1em}{0.8em}{\rotatebox{90}{\text{Register 2}}}}
}
\end{equation*}
\caption{The generalized circuit for the quantum subset-sum algorithm}
\label{generalized_circuit}
\end{figure}

Representing the action of the QPE algorithm by the \textit{gate} $U_{\text{QPE}}$, which includes the QFT$^{-1}$ operations, we arrive at a state $\ket{\psi_1}$ shown as the dashed box on the left side of Fig. \ref{generalized_circuit},

\begin{equation}
        \ket{\psi_1} = U_{\text{QPE}} \ket{\psi_0} = \frac{1}{\sqrt{2^n}} \sum_{j=0}^{2^n-1} \ket{\varphi_j} \ket{j}. \label{psi1}
\end{equation}

Note that the simplicity of our unitary matrix $U$, namely that it is easily implemented by IBM Quantum Experience default $T$ gates. Furthermore, the QFT$^{-1}$ operation is $\mathcal{O}(t^2) = \mathcal{O}(\text{poly}(n))$. The complexity to reach the state in Eq. \ref{psi1} is $\mathcal{O}(\text{poly}(n))$.

\subsection{Eliminating The \textit{Bad} Subsets}

Following the QPE procedure we are left with a state containing the superpositions of all possible subset-sums and the associated elements for each sum stored in the registers. We now wish to apply Amplitude Amplification (AA) to find our solution. To begin this next phase of the quantum computation, start by formally splitting the state $\ket{\psi_1}$ into two parts,

\[ \ket{\psi_1} = \frac{1}{\sqrt{2^n}} \sum_{j \in L} \ket{\varphi_j} \ket{j} + \frac{1}{\sqrt{2^n}} \sum_{j \in L'} \ket{\varphi_j} \ket{j} = \sqrt{\frac{|L|}{2^n}} \ket{\psi_{\text{good}}} + \sqrt{\frac{|L'|}{2^n}} \ket{\psi_{\text{bad}}}, \] 
where $L = \{j : \varphi_j \leq s\}$ and $L' = \{j : \varphi_j > s\}$ with $0 \leq j < 2^n$.

The AA procedure as discussed in \cite{NielsenChuang} is now applied using the iterator $G = S (F_{\varphi} \otimes I^{\otimes n})$, where $F_{\varphi}$ operates on Register 1 and flips the sign of any state with $\varphi_j \leq s$,

\[ (F_{\varphi} \otimes I^{\otimes n}) \ket{\psi_1} = -\sqrt{\frac{|L|}{2^n}} \ket{\psi_{\text{good}}} + \sqrt{\frac{|L'|}{2^n}} \ket{\psi_{\text{bad}}}. \]

For this case it is easy to show that $S = 2\ket{\psi_1}\bra{\psi_1} - I$ and is implemented using the gates

\[ S = (I^{\otimes t} \otimes H^{\otimes n})U_{\text{QPE}}U_{0\perp}(I^{\otimes t} \otimes H^{\otimes n})U_{\text{QPE}}^*, \]
        where $U_{0\perp} = I - 2\ket{\mathbf{0}}\bra{\mathbf{0}}$.

        The implementation of $F_\varphi$ involves a simple combination of $Z$ and $X$ gates. The known value of the desired sum $s$ can be loaded as the phase value of the multiplier in the $Z$ gate. Heuristically, applying $X$ and then $Z$ to a trial state as part of the \textit{oracle} lookup first flips that state's phase and then adds the true value to that phase. For states nearly close to the desired sum, this leaves a state with nearly negligible phase amplitude and this state is thus \textit{marked} as a known good state. If \textit{good}, then $Z^\dagger$ is applied to reverse the phase change, leaving only $X$ applied in order to flip that state's bit.  If the state was a \textit{bad} candidate, $X^\dagger Z^\dagger$ is applied to return to the original state.

        Repeated iterations of $G$ (dotted box in Fig. \ref{generalized_circuit}) amplify the \textit{good} states and work to eliminate the \textit{bad} states. The number of such necessary iterations is bounded by $\mathcal{O}(\sqrt{\frac{2^n}{{|L|}}})$. If we assume that $\frac{|L'|}{|L|} = \mathcal{O}(\text{poly}(n))$, then the number of these iterations is bounded in time by $\mathcal{O}(\text{poly}(n))$.

At the end of the AA phase we are left with the state

\begin{equation}
        \ket{\psi_2} \approx \frac{1}{\sqrt{|L|}} \sum_{j \in L} \ket{\varphi_j} \ket{j}. \label{psi2}
\end{equation}

This is the input to the final computation phase, the second dashed box in Fig. \ref{generalized_circuit} where we next find the maximum value among all the subset-sums.

\subsection{Finding The Maximum Subset-sum}
We could make use of Grover's search algorithm to find the maximum value of $\varphi_j$ using $\ket{\psi_2}$ in $\mathcal{O}(\sqrt{|L|})$ times but because of the assumption made above this would lead to an overall exponential running time for the entire computation. Instead, we make use of the fact that the elements of the set $\{\varphi_j : 0 \leq j < 2^n\}$ are partially sorted and for the most part $\varphi_j \leq \varphi_{j+r}$ for relatively large $r$. We can then use a quicker binary search algorithm to find the solution in $\mathcal{O}(\text{large} |L|) = \mathcal{O}(\text{poly}(n))$ time. The procedure below follows the same logic as the verification procedure in \cite{Marriott2005}.

We apply a series of conditional amplitude amplifications. Assume the maximum $\varphi_j$ in Eq. \ref{psi2} is $\varphi_{\text{max}} = (b_0 \ldots b_{t-1})_2$. Attempting to maximize the measurement outcome of the first qubit of Register 1, we obtain a value close to $\varphi_{\text{max}}$. This process is done by measuring the most significant qubits, trying to record as many $\ket{\mathbf{1}}$ states as possible. In algorithmic form, then,

\begin{enumerate}
        \item Measure the most significant qubit.
        \item If the outcome is not $\ket{\mathbf{1}}$, apply AA to amplify the states where this qubit is in state $\ket{\mathbf{0}}$ and then measure it again. If this qubit does not yield $\ket{\mathbf{1}}$ within a few iterations, then we assume $\ket{\mathbf{0}}$ is the value of this qubit and move on to the next qubit.
\end{enumerate}

This procedure is illustrated as the right-hand dashed box in Fig. \ref{generalized_circuit}. The $S_1, S_2, \ldots, S_t$ are virtually identical to the original $S$ but contain the total sum of all quantum operations up to the specific invocations. That is,

\[ S_1 = U_1U_{0\perp}U_1^*, \]
where the $U_1$ represents all quantum operations done up to the point where $S_1$ is applied. The generators of the AA procedure in this stage are also given by

\[ G_1 = S_1 (Z \otimes I^{\otimes tn-1}), G_2 = S_2 (Z \otimes I^{\otimes tn-2}), \ldots, \]
and each $G_i$ is repeated until a measurement of that qubit results in a $\ket{\mathbf{1}}$. If it does not, $\ket{b_i}$ is set to $\ket{0}$ and we move on.

We are making a second assumption for this final procedure to work, namely that if the bit value of any $i^{th}$ qubit is $1$ in the binary value of $\varphi_{\text{max}} = (b_0 \ldots b_t-1)_2$; then after measuring the first $(i-1)$ qubits with the correct values $(b_0 b_1 \ldots b_{i-2})_2$, the probability of seeing a $\ket{1}$ on the $i^{th}$ qubit is not exponentially small in the normalized collapsed state. Note that this assumption does not affect the overall $\mathcal{O}(\text{poly}(n))$ running time of our algorithm but rather it only effects the accuracy of our result as we only run the repeated AA steps toward the final solution a small number of times. A few runnings of the actual algorithm may be required to achieve a correct results. Thus, the total runtime complexity of the entire algorithm should be $\mathcal{O}(\text{poly}(n))$. 

\section{Summary}

We have presented a new quantum algorithm that solves the Subset-sum problem for certain not-too-hard cases in $\mathcal{O}(\text{poly}(n))$ time. However, the required number of qubits in the two registers used to implement the solution make this difficult to implement on the current size of the IBM Quantum Experience machine available to us. Furthermore, due to lack of time on the part of the authors, there was no time to develop even the smallest case example to test out the procedures and complexity analysis outlined in this note. We look forward to furthering this work given a future opportunity.

\section{Acknowledgment}
Los Alamos National Laboratory, an affirmative action/equal opportunity employer, is operated by Los Alamos National Security, LLC, for the National Nuclear Security Administration of the U.S. Department of Energy under contract DE-AC52-06NA25396.

\end{document}